\documentclass[10pt,conference]{IEEEtran}
\IEEEoverridecommandlockouts
\usepackage{cite}
\usepackage{amsmath,amssymb,amsfonts}
\usepackage{algorithmic}
\usepackage{algorithm}
\usepackage{mathtools}
\usepackage{graphicx}
\usepackage{textcomp}
\usepackage{xcolor}
\usepackage{makecell}
\usepackage{subcaption}
\usepackage{multirow}
\usepackage{dblfloatfix}
\def\BibTeX{{\rm B\kern-.05em{\sc i\kern-.025em b}\kern-.08em
    T\kern-.1667em\lower.7ex\hbox{E}\kern-.125emX}}

\IEEEoverridecommandlockouts \IEEEpubid{\makebox[\columnwidth]{978-1-6654-0601-7/22\/\$31.00 ~\copyright~2022 IEEE \hfill} \hspace{\columnsep}\makebox[\columnwidth]{ }}

\begin{document}

\newcommand{\system}{ROMA}

\title{\system: Resource Orchestration for Microservices-based 5G Applications
{\footnotesize \textsuperscript{}}

}
\makeatletter
\newcommand{\linebreakand}{%
  \end{@IEEEauthorhalign}
  \hfill\mbox{}\par
  \mbox{}\hfill\begin{@IEEEauthorhalign}
}
\makeatother
\author{\IEEEauthorblockN{1\textsuperscript{st} Anousheh Gholami \textsuperscript{*} \thanks{\textsuperscript{*} Work done as an intern at NEC Laboratories America, Inc.}}
\IEEEauthorblockA{
\textit{University of Maryland} \\
College Park, MD \\
anousheh@umd.edu}
\and
\IEEEauthorblockN{2\textsuperscript{nd} Kunal Rao}
\IEEEauthorblockA{
\textit{NEC Laboratories America}\\
Princeton, NJ \\
kunal@nec-labs.com}
\and
\IEEEauthorblockN{3\textsuperscript{rd} Wang-Pin Hsiung}
\IEEEauthorblockA{
\textit{NEC Laboratories America}\\
San Jose, CA \\
whsiung@nec-labs.com}
\and
\IEEEauthorblockN{4\textsuperscript{th} Oliver Po}
\IEEEauthorblockA{
\textit{NEC Laboratories America}\\
San Jose, CA \\
oliver@nec-labs.com}
\linebreakand
\IEEEauthorblockN{5\textsuperscript{th} Murugan Sankaradas}
\IEEEauthorblockA{
\textit{NEC Laboratories America}\\
Princeton, NJ \\
murugs@nec-labs.com}
\and
\IEEEauthorblockN{6\textsuperscript{th} Srimat Chakradhar}
\IEEEauthorblockA{
\textit{NEC Laboratories America}\\
Princeton, NJ \\
chak@nec-labs.com}
}



\maketitle

\begin{abstract}
With the growth of 5G, Internet of Things (IoT), edge computing and cloud computing technologies, the infrastructure (compute and network) available to emerging applications (AR/VR, autonomous driving, industry 4.0, etc.) has become quite complex. There are multiple tiers of computing (IoT devices, near edge, far edge, cloud, etc.) that are connected with different types of networking technologies (LAN, LTE, 5G, MAN, WAN, etc.). Deployment and management of applications in such an environment is quite challenging. In this paper, we propose \system, which performs resource orchestration for microservices-based 5G applications in a dynamic, heterogeneous, multi-tiered compute and network fabric. We assume that only application-level requirements are known, and the detailed requirements of the individual microservices in the application are not specified. As part of our solution, \system\ identifies and leverages the coupling relationship between compute and network usage for various microservices and solves an optimization problem in order to appropriately identify how each microservice should be deployed in the complex,  multi-tiered compute and network fabric, so that the end-to-end application requirements are optimally met. We implemented two real-world 5G applications in video surveillance and intelligent transportation system (ITS) domains. Through extensive experiments, we show that \system\ is able to save up to $90\%$, $55\%$ and $44\%$ compute and up to $80\%$, $95\%$ and $75\%$ network bandwidth for the surveillance (watchlist) and transportation application (person and car detection), respectively. This improvement is achieved while honoring the application performance requirements, and it is over an alternative scheme that employs a static and overprovisioned resource allocation strategy by ignoring the resource coupling relationships.
\end{abstract}

\begin{IEEEkeywords}
resource orchestration, IoT, 5G, edge computing, microservices, system modelling and optimization 
\end{IEEEkeywords}

\section{Introduction}
\label{intro}
The fifth generation (5G) of mobile network promises to support a wide range of IoT services requesting strict and diverse set of requirements in terms of end-to-end latency, throughput, reliability, etc. In order to accommodate these services automatically and at large scale, network slicing has emerged as an evolutionary solution for the design and deployment of the next generation mobile networks, enabled by the programmability and flexibility that software defined networking (SDN) and network function virtualization (NFV) technologies introduce into the future network management systems. A network slice (NS), in the context of 5G, is composed of sub-slices encompassing the radio access network (RAN), core network (CN) and the transport network. 
3GPP has put efforts into integrating network slicing in the future specification of both RAN and CN domains\cite{3gpp}. Despite significant benefits that network slicing has demonstrated to bring into the mobile network systems management and performance, the real-time response required by the delay-sensitive applications, such as autonomous driving, video analytics and streaming applications, necessitates the integration of the multi-access edge computing (MEC) into 5G networks and beyond. The aim of MEC is to push different resources from the remote central cloud to the network edge in close vicinity of users and IoT sensors, where data is generated \cite{mec}. 

\begin{figure}[t]
    \centering
    \includegraphics[width=0.65\linewidth]{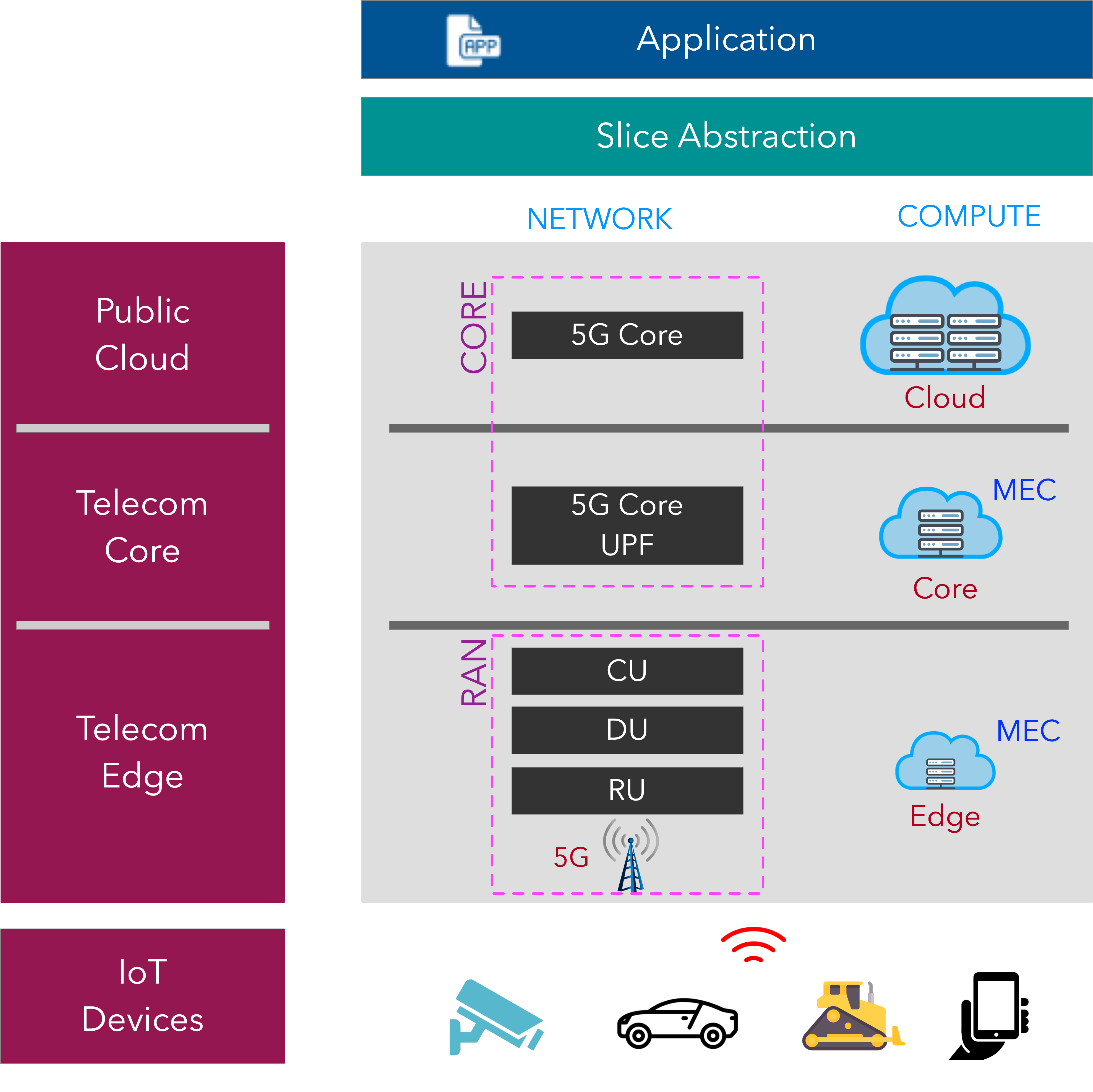}
    \caption{Multi-tiered compute and network fabric}
    \label{fig-architecture}
\end{figure}

Prior to MEC, mobile cloud computing (MCC) became a key technology to realize computationally-intensive applications by offloading substantial amounts of mobile devices' computing functionality to a remote cloud data center \cite{fernando2013mobile}. Compared to MCC, MEC has lower latency, but it can easily become overloaded. Therefore, a multi-tiered computing and networking system where critical services are offloaded to MEC and the delay-tolerant services are computed at the remote cloud has the potential to improve the applications performance and the overall resource utilization \cite{collaborative}. Fig. \ref{fig-architecture} shows such a multi-tiered compute and network fabric. We note that there is a slice abstraction on top of the compute and network infrastructure and application uses this slice abstraction to request network as well as compute slices. Moreover, the underlying infrastructure components are untouched by the application. All requests for compute and network slices always go through slice abstraction, which may grant or deny requested slices depending on the resource conditions and demands at that time. In such a multi-tiered architecture, compute is available at various tiers like devices i.e. where data is produced, edge and in the cloud. Similarly, different kinds of networking capabilities are available at different tiers, e.g. 5G connectivity between devices and edge servers, MAN between distributed edge resources and WAN between edge and central cloud. Deploying and optimizing applications in such a complex infrastructure is very challenging.
Moreover, the real-time state of different resources (e.g. available network and compute) is highly dynamic due to high variability in the compute (multi-tenancy, heterogeneity, etc.), and changing network (5G NR interference, link congestion, packet loss, etc.) conditions. Therefore, the problem of application deployment and optimization across multi-tiered compute and network fabric is even more challenging considering the real-time state of resources.

For an application which is composed of multiple individual microservices a.k.a. functions, deployment of the application entails  (i) deployment of individual functions, and (ii) management of data communication between various functions. Fig. \ref{fig:watchlist service} shows the pipeline of such a microservices-based watchlist application, for video surveillance use cases. Each individual function for this application has demands for different resources such as compute, storage, bandwidth for the incoming streams, etc. The overall performance of the application depends on the allocated resources to all the microservices. This introduces a coupling relationship between the usage of different resources and the application performance, which we present in detail in section \ref{sec:motivation}. To the best of our knowledge, this is the first work which studies and incorporates these coupling relationships in order to optimally allocate resources to the functions of microservices-based 5G applications.
\begin{figure}[t]
    \centering
    \includegraphics[width=0.95\linewidth]{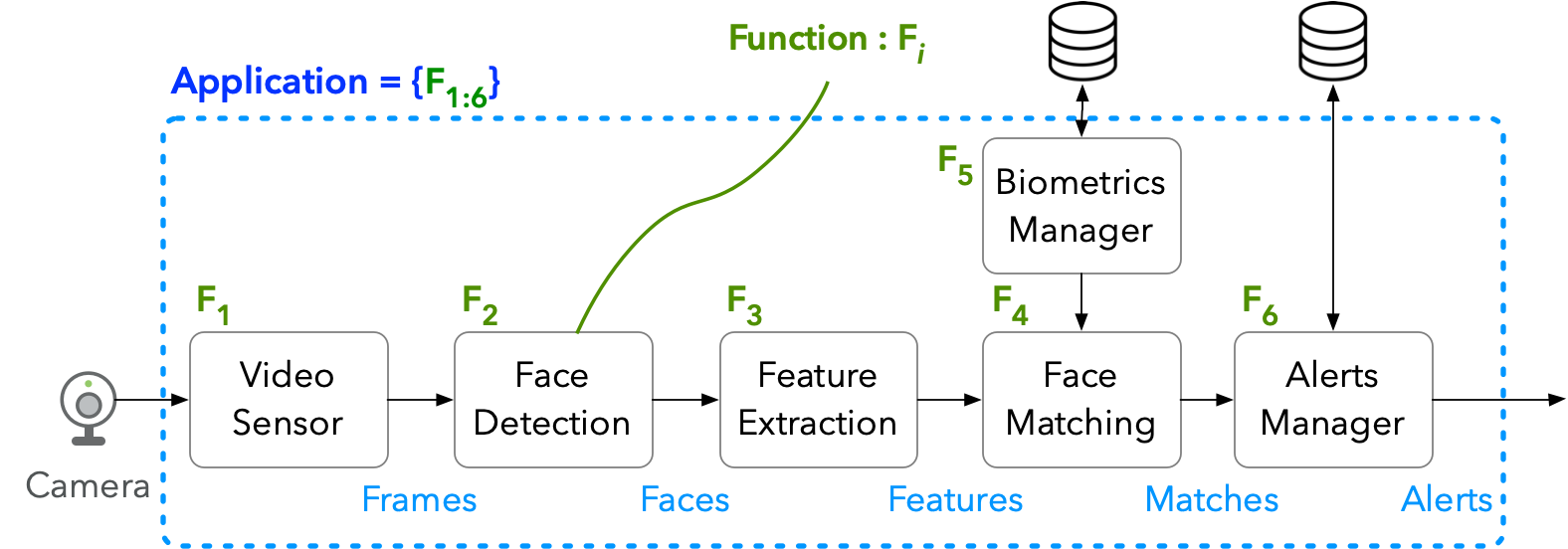}
    \caption{Video surveillance: watchlist application}
    \label{fig:watchlist service}
\end{figure}
\begin{figure}[t]
    \centering
    \includegraphics[width=0.68\linewidth]{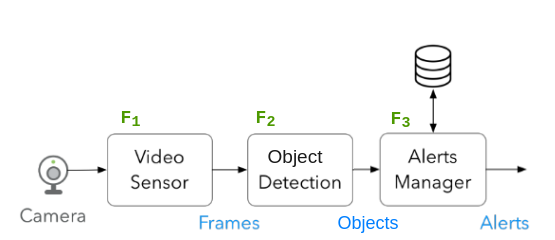}
    \caption{Intelligent transportation system: object detection application}
    \label{fig:object detecting service}
\end{figure}

\begin{figure}[t]
    \centering
    \includegraphics[width=0.90\linewidth]{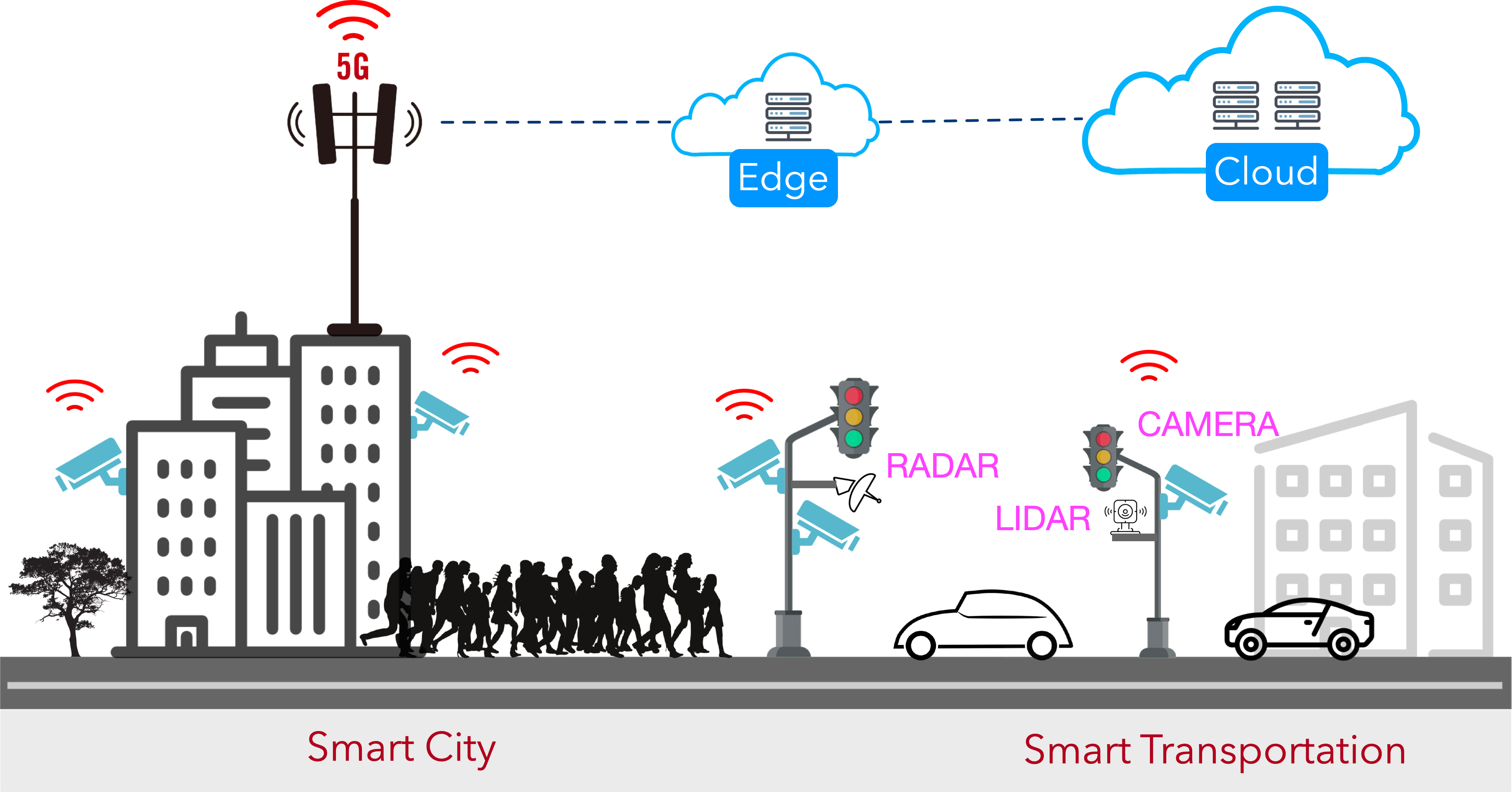}
    \caption{City-scale IoT sensors deployment}
    \label{fig-city-scale}
\end{figure}

\begin{figure*}[t]
\centering
  \begin{subfigure}{0.32\textwidth}
    \includegraphics[width=\linewidth]{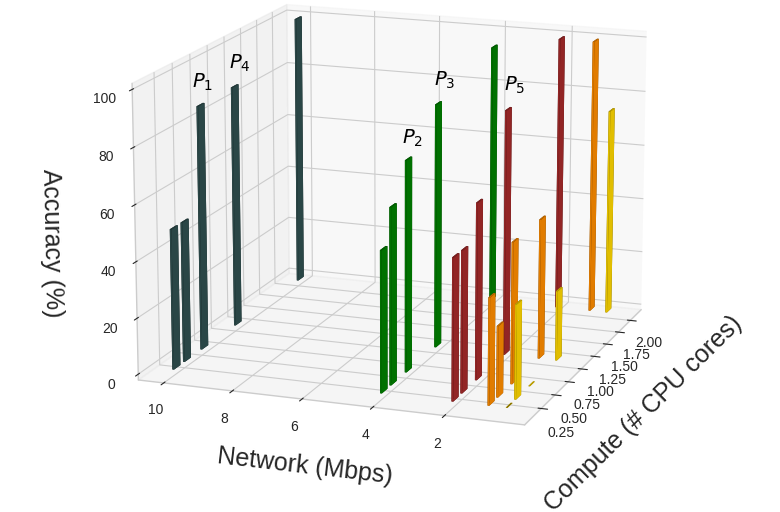}
    \caption{Exp 1-1 - Watchlist}
    \label{fig:watchlist-com-br-lab}
  \end{subfigure}
  \begin{subfigure}{0.32\textwidth}
    \includegraphics[width=\linewidth]{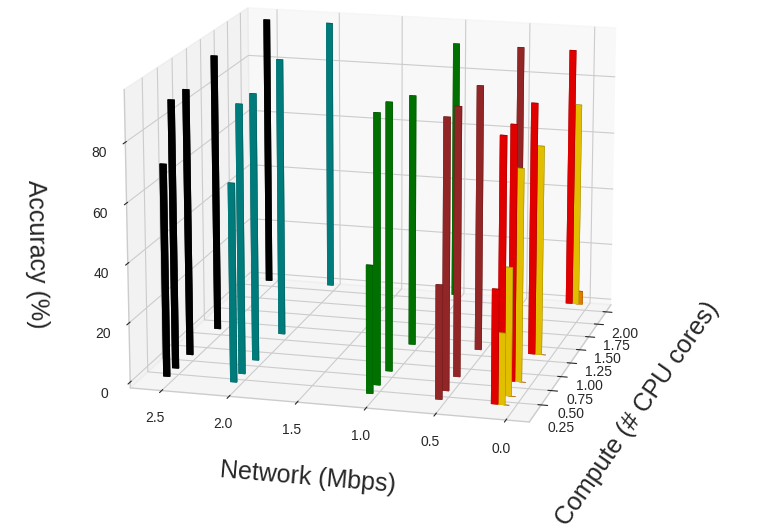}
    \caption{Exp 1-2 - Watchlist}
    \label{fig:watchlist-com-br-pol}
  \end{subfigure}\\
  \begin{subfigure}{0.32\textwidth}
    \includegraphics[width=\linewidth]{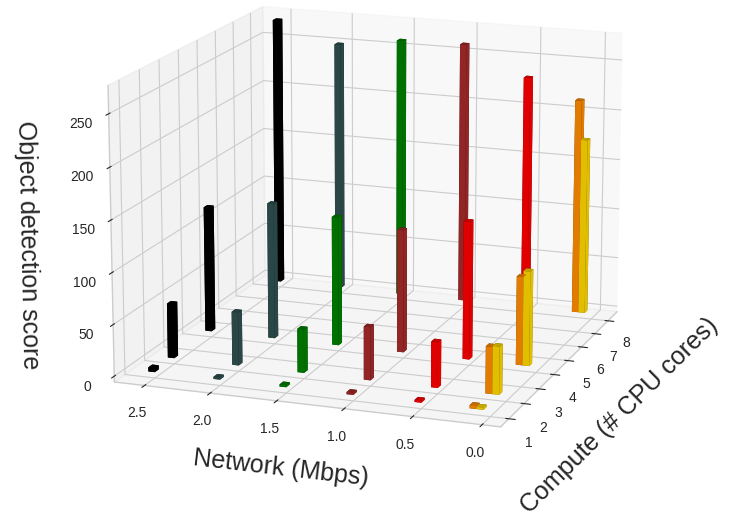}
    \caption{Exp 2-1 - Object (person) detection}
    \label{fig:com-br-pol}
  \end{subfigure}
  \hspace*{\fill}
  \begin{subfigure}{0.32\textwidth}
    \includegraphics[width=\linewidth]{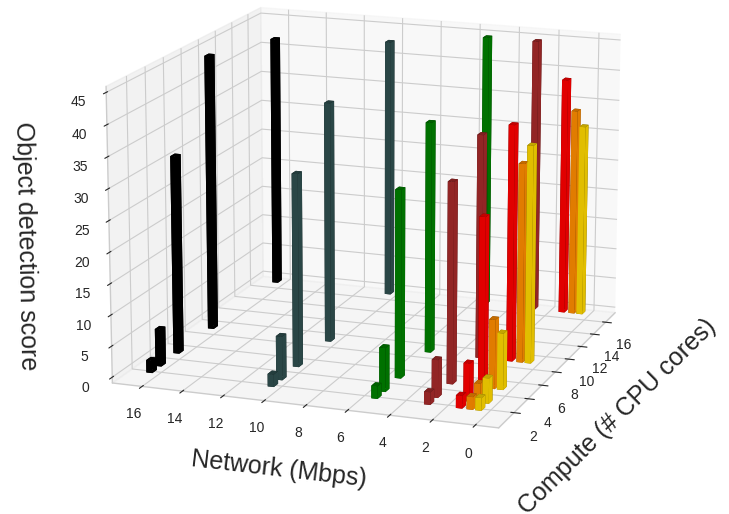}
    \caption{Exp 2-2 - Object (person) detection}
    \label{fig:com-br-lab}
  \end{subfigure}
    \hspace*{\fill}  
  \begin{subfigure}{0.32\textwidth}
    \includegraphics[width=\linewidth]{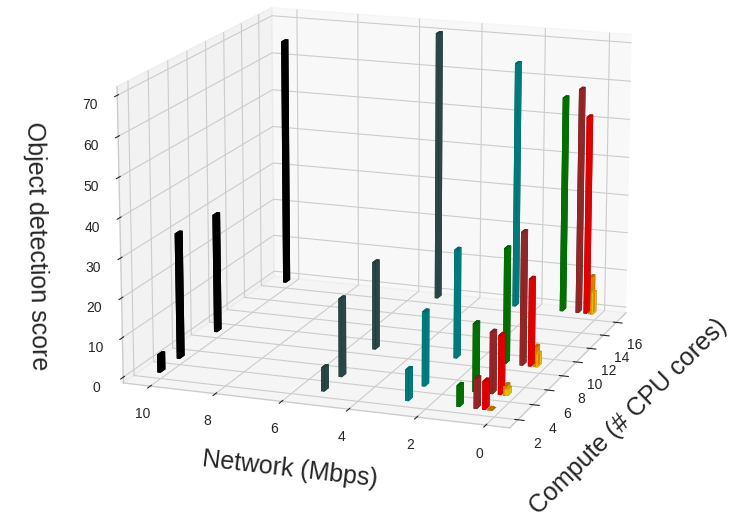}
    \caption{Exp 2-3 - Object (car) detection}
    \label{fig:com-br-car}
  \end{subfigure}
\caption{Impact of network and compute usage on the application performance} \label{fig:coupling}
\end{figure*}
A multi-tiered infrastructure opens up several possibilities for the deployment of functions and inter-connections between them. Thus, the questions of where (e.g. which compute node or tier) each individual function is deployed at and how the communication between functions is realized, taking into account the resource coupling relationships, are open research problems.
The placement decisions directly impact the bottom-line for the application e.g. end-to-end application latency, accuracy, cost of resource usage or any other application-level metric. Therefore, proper function-level resource allocation and automatically adjusting function-level resources in response to changes in the dynamic infrastructure is critical for meeting application-level requirements. The resource allocation problem becomes even more severe at a large-scale deployment e.g. city-scale as shown in Fig. \ref{fig-city-scale}, when there are hundreds to thousands of IoT sensors, each continuously producing data stream, which needs to be transmitted over 5G for local/remote processing. In such scenarios, managing function-level resources for processing all these data streams in a dynamic environment is a very challenging task.

The goal of this paper is to automatically manage the execution of microservices-based application on a dynamic, heterogeneous, multi-tiered compute fabric in a 5G network taking into account the coupling relationships between resources, when only application-level requirements are given, without knowing individual function-level requirements. The following are the main contributions of this paper:
\begin{itemize}
    \item We first identify and model the coupling between network and compute resource usage, and analyze the impact of the coupling on application performance by considering multiple real-world use cases.
    \item Then, we propose a novel optimization formulation that captures the compute and network coupling relationship, and enables a principled consideration of resource allocation options to significantly reduce network and compute resource usage.  Our proposed runtime system (referred to as \system) uses the new formulation, and utilizes the coupling to jointly optimize compute and network resources across different edge compute and network slices while ensuring consistent, high quality insights.
    \item We implement two real-world IoT applications in video surveillance and intelligent transportation system domains and show that \system\ can improve resource usage significantly (save up to $90\%$ compute resource usage, and up to $95\%$ network resource usage), while maintaining quality of insights, compared to a static resource allocation approach that ignores compute and network resource usage coupling.
\end{itemize}
The paper is organized as follows. In section \ref{sec:motivation}, we elaborate on the main motivation of this paper and describe the challenges in \ref{sec:challenges}. Section \ref{sec:system model} describes the system model. In section \ref{sec:solution}, we introduce our solution method. Performance evaluation is presented in section \ref{sec:evaluation}, while we provide the literature review in section \ref{sec:related}. Finally, in section \ref{sec:conclusion}, we highlight our conclusions.
\section{Motivation}
\label{sec:motivation}
In this section, we highlight the motivation behind the addressed problem through numerical experiments. We consider two video analytics use cases in two different industry verticals. The first one is in video surveillance and the second one in ITS. Fig. \ref{fig:watchlist service} shows the structure of the video surveillance use case i.e. watchlist application, which uses face recognition technology to identify individuals seen in front of a camera. Fig. \ref{fig:object detecting service} shows object detection application, which is used in ITS to detect objects like cars and even people e.g. pedestrians, and build higher-level applications like accident prevention, safety alerting, traffic control, etc.

Fig. \ref{fig:coupling} illustrates the impact of the network and compute resources on the performance of the above applications. In Fig. \eqref{fig:watchlist-com-br-lab} and \eqref{fig:watchlist-com-br-pol}, the performance of the watchlist application in terms of face detection accuracy is evaluated for two sample videos as the allocated CPU cores to the face-detection function and the network bandwidth for the input streams of the video sensor ingress function vary. Fig. \eqref{fig:com-br-pol} and \eqref{fig:com-br-lab} show the performance of the object (person) detection application in terms of the detection score (defined in section \ref{sec:evaluation}) for two sample videos. In Fig. \eqref{fig:com-br-car}, the performance of the object detection application for a car detection task is illustrated for different compute and network availabilities. It is important to note that in all cases, the performance of the application is controlled by both network and compute resource usage. As a result, in order to  avoid resource overprovisioning and meet application requirements, this coupling effect should be considered when deploying the application. From Fig. \eqref{fig:coupling}, it is observed that although the general pattern is similar in different experiments (increasing compute and network improves the performance in most of the cases), the coupling relationship is application-specific and even within an application, it is non-linear and therefore not trivial. 

In the following, we clarify the coupling relationships and how they can be utilized to optimize the resource allocation decision or enhance the application performance through three specific operation instances of experiment 1-1 (Fig. \eqref{fig:watchlist-com-br-lab}). Suppose that initially, the watchlist application is operating at point $P_1=(0.5\ core, 10\ Mbps)$. If the network experiences congestion (network is a bottleneck) and the bandwidth drops from $10\ Mbps$ to $4\ Mbps$, the system is forced to operate at point $P_2 = (0.5\ core, 4 \ Mbps)$, and the performance of the application drops from $83\% $ to $71\%$. In this scenario, if there exist idle compute, the performance can be improved by increasing the allocated CPU cores to $1\ core $ and moving to $P_3 = (1\ core, 4\ Mbps)$. In an inverse order, if the system is initially deployed at $P_3$ with performance of $83\%$ and suddenly the available compute resource reduces (compute is a bottleneck), the performance degrades by $12\%$ by moving from $P_3$ to $P_2$. In this case, allocating more network bandwidth to the incoming video stream changes the operating point to $P_1$, thus performance remains unchanged. In each of the aforementioned cases, there is a tradeoff between network and compute usage which can be exploited to avoid performance degradation by taking into account resource coupling relationships in the application resource orchestration and devising a joint network and compute resource allocation scheme.

Now consider a third case in which the system is operating at point $P_4=$ $(1\ core, 10\ Mbps)$ initially. It is observed that the allocated compute resource can be reduced to $0.5\ core$ by moving to $P_1 = (0.5\ core, 10\ Mbps)$ without affecting the performance, thus avoiding resource overprovisioning. The released CPU cores can be allocated to other services deployed on the same node, which enhances the resource utilization and the performance of other deployed applications. Another option is to reduce the network usage form $10\ Mbps$ to $4\ Mbps$ and reduce even more to $2\ Mbps$ by moving to point $P_5 = (1 \ core, 2 \ Mbps)$, without affecting the application accuracy. Further reduction of network bandwidth will result in significant accuracy degradation and thus should be avoided. Therefore, even though neither network nor compute resources become scarce, there may exist multiple paths or options to save on different resources and avoid overprovisioning while keeping the performance unchanged or within an acceptable range. The decision on which path (operational point) to be opted depends on the objective of the application manager and the state of the resources. In the following, we propose an optimization based decision making process for application deployment incorporating the resource coupling relationships. 

\section{Challenges}
\label{sec:challenges}
In this section, we discuss the main challenges we faced while designing \system\ and the approach we propose to address them.\\
\textit{Challenge 1: Identifying the resource coupling relationship and its impact on the application performance}. The performance of a microservices-based application is controlled by the amount of different resources used by different microservices as discussed in section \ref{sec:motivation}. These coupling relationships are not only application-specific, but also time-variant. Ignoring this important phenomenon in the application resource orchestration phase may result in huge resource overprovisioning and undesirable performance. While the resource coupling relationships can be non-linear in general, in this paper, we use linear regression to model the application performance as a function of the allocated resources and illustrate the effectiveness of this approach through extensive experiments.
\\\\
\textit{Challenge 2: Unknown function-level performance requirements}. In a multi-tiered 5G infrastructure, the decisions on the placement of application functions and the resource allocations determine the application-level performance metrics such as end-to-end latency, throughput, accuracy, etc. While knowing the functions-level requirements simplifies the deployment process, in a realistic setup, only application-level requirements are known. In this paper, we propose an optimization model with function placement and resource allocation decision variables, and the application-level performance requirements are modeled as hard constraints.
\section{System Model}
\label{sec:system model}
In the sequel, we present the system model and the problem formulation. The physical infrastructure consists of computing nodes distributed across multiple layers, at the edge and at a central cloud. At each compute tier, compute slicing is possible for the allocation of the resources to different applications. 
Let $\mathcal{M}$ represent the set of compute nodes. Each compute node $m\in \mathcal{M}$ is specified by $(\boldsymbol{g}_m, tier_m)$, where $\boldsymbol{g}_m$ and $tier_m$ denote the available resource vector and the associated tier (e.g. IoT device, far edge, near edge, central cloud), respectively.
Assuming that each node $m$ provides $T$ different resources represented by set $\mathcal{T}$, the size of $\boldsymbol{g}_m$ is $T$. While our proposed approach can be easily extended to arbitrary set of resources, in this paper, we consider network and compute resources, i.e. $\mathcal{T} = \{com, net\}$.

We model an application as a set of microservices or functions and interconnections that represent the data dependency between functions. An application is specified by a tuple $\mathcal{R}= (\tau, \omega)$, where $\tau$ and $ \omega$  stand for the required end-to-end delay and throughput of the application, respectively. 
Let $G=(V,E)$ be the graph representing the application, where $V$ denotes the set of application functions and $E$ represents the interconnections between functions. Furthermore, $\mathcal{R}_v = (\tau_v, \omega_v)$ denotes the portion of the delay and throughput corresponding to node (function) $v$. Moreover, $tier^v$ denotes the tier on which function $v$ should run, if such constraints exists for function $v$. For instance, there might exist constraints on some functions of the mobile application (such as in user-initiated applications) to run locally (on the user equipment).
Given the function level performance metrics $\{\mathcal{R}_v, v\in V\}$, we assume that the rules defining the application level performance metrics are known. One challenge in this regard is to determine the set of functions contributing to each of the end-to-end application performance metrics (a.k.a the critical path or pipeline of the application). For the sake of simplicity, we assume that the knowledge about the contributing functions to each performance metric is available by the application developer similar to \cite{ECO}. For instance, given the functions of the critical path of $G$ as $V_{critical} \in V$, we can calculate the end-to-end application delay as $h_{delay}(\tau_1,...,\tau_{|V|}) = \sum_{v\in V_{critical}} \tau_v$. Similarly, the throughput rule is computed as $h_{throughput}(\omega_1,...,\omega_{|V|})= min_{v\in V_{critical}} \omega_v$.

In order to successfully and optimally deploy an application given its end-to-end performance requirements, it is important to understand the coupling between the usage of different resources. Let $p$ denote the desired application performance, e.g. $p$ can be the detection accuracy in the watchlist application. To address the impact of the network and compute resources on the application performance, we define $f^{t,t^\prime}_{v,v^{\prime}} (x,p):\mathbb{R}\xrightarrow{} \mathbb{R}$, $v,v^{\prime} \in V$, $t, t^\prime \in \mathcal{T}$ as the minimum resource unit of type $t^\prime$ that should be allocated to function $v^{\prime}$ in order to achieve the application performance of $p$, given that $x$ units of resource type $t$ is allocated to function $v$. In fact, $f^{t,t^\prime}_{v,v^{\prime}}(.,.)$ reflects the coupling relationship between each pair of resources allocated to all pairs of application functions. 
Even for the same function i.e. when $v= v^{\prime}$, coupling relationship between different types of resources is reflected through the defined function as well. For instance, given that the input streams of function $F_2$ of the watchlist application in Fig. \ref{fig:watchlist service} consume $ x \ Mbps$ network, the minimum number of CPU cores that should be allocated to $F_2$ in order to achieve an accuracy of $p $ is equal to $f^{net,com}_{2,2}(x,p)$. As a numerical example, for the watchlist application of Fig. \ref{fig:watchlist service}, it can be observed from Fig. \eqref{fig:watchlist-com-br-lab} that $f^{net,com}_{1,2}(10, 80) = 0.75$, i.e. in order to have the accuracy of $80\%$ when the available network bandwidth for the video stream input of $F_1$ is $10\ Mbps$, it suffices to allocate $0.75$ CPU core to the face detection function $F_2$.

\section{Problem Formulation and Proposed Solution}
\label{sec:solution}
We model the application deployment problem across a multi-tiered compute and network fabric as an optimization problem. We then discuss the usage of different models for the coupling relationships.

\subsection{Optimizing resource allocation and application performance}
\label{optmodel}
The application resource allocation and performance optimization problem entails the assignment of microservices to the compute nodes in $\mathcal{M}$ (a.k.a. placement problem) and the allocation of different resources to each function, such that end-to-end  application requirements (e.g. delay and throughput) are satisfied. We model this problem as a  multi-objective optimization problem, with the objective of minimizing the total resource usage (equivalently, the deployment cost) and maximizing the application performance, by incorporating the resource coupling functions introduced in section \ref{sec:system model}. By designing a joint optimization problem with two objective terms, the tradeoff between performance and resource usage illustrated in the examples of section \ref{sec:motivation} is also captured. The following decision variables are defined for the problem formulation:
\begin{itemize}
\item $x_{v,m}$: a binary decision variable for function placement which is equal to $1$ if the function $v$ of the application is assigned to the substrate node $m$ for execution and $0$ otherwise.
\item $y^{t}_{v,m}$: a continuous decision variable denoting the amount of resource type $t$ of node $m$ allocated to function $v$.
\item $p$: a continuous decision variable representing the application performance, e.g. the face recognition accuracy or object detection score.
\end{itemize}
The resulting optimization problem is as follows:

\begin{align}
    & [P] \quad \textbf{min} \quad \eta \sum_{t,v,m}{y^{t}_{v,m}} - (1-\eta) p \label{opt:objective}\\
    &\textbf{s.t.}\nonumber \\  
    &\sum_{m} x_{v,m} f^{t,t^\prime}_{v,v^{\prime}} (y^{t}_{v,m},p) \leq  \sum_{m} x_{v^\prime,m} y^{t^\prime}_{v^\prime,m}, \forall v,v^\prime, t,t^\prime \label{opt:rec1}\\
    & y^t_{v,m}\leq g^t_m x_{v,m}, \forall t,m,v \label{opt:rec2-2}\\
    & \sum_{v\in V} y^t_{v,m}\leq g^t_m, \forall t,m \label{opt:rec2}\\
    &\sum_{m|tier^v = tier_m}  x_{v,m}=1, \forall v \label{opt:map1}\\
    & \tau \geq h_{delay}(\tau_1,...,\tau_{|V|}) \label{opt:perf1}\\
    & \omega \leq h_{throughput}(\omega_1,...,\omega_{|V|})\label{opt:perf2}\\
    & x_{v,m} \in \{0,1\}, \forall v\in V, m\in\mathcal{M} \nonumber\\
    & 0 \leq y^{t}_{v,m} \leq g^t_m , \forall v\in V, m\in\mathcal{M}, t\in \mathcal{T}, \ 0 \leq p \leq p_{max} \label{opt:dom}
\end{align}
In the objective function \eqref{opt:objective}, $\eta$ is a parameter between $0$ and $1$ used to control the balance between the two objective terms. In our experiments, we tested different values for $\eta$ and selected a small value to promote a solution that primarily enhances the performance and minimizes the total consumed resources. Constraints \eqref{opt:rec1} ensure that the resources allocated to each application microservice is greater than or equal to the required minimum amount (given by the defined coupling functions) to potentially achieve the performance of $p$. For instance, the resource type $t^\prime$ allocated to function $v^\prime$ which is equal to $\sum_m x_{v^\prime, m} y^{t^\prime}_{v^\prime,m}$ should be greater than or equal to $\sum_{m} x_{v,m} f^{t,t^\prime}_{v,v^{\prime}} (y^{t}_{v,m},p)$ for all $t,v$. This constraint together with the objective of minimizing total used resources results in a solution which avoids resource overprovisioning. The set of inequalities in \eqref{opt:rec2-2} and \eqref{opt:rec2} enforce the infrastructure capacity constraints. Constraints \eqref{opt:map1} ensure that each function of an application is deployed at one infrastructure node. The application end-to-end performance requirements are guaranteed by constraints \eqref{opt:perf1} and \eqref{opt:perf2}. Finally, the domain constraints are expressed in \eqref{opt:dom}, where $p_{max}$ is the maximum observed performance for a specific application in all resource allocation vectors. The optimization problem $[P]$ is a mixed integer nonlinear program (MINLP) owning to the constraints \eqref{opt:rec1} and the integer variables, thus an $\mathcal{N}\mathcal{P}$-hard problem. In the next section, we discuss the models for the coupling functions and solve a special case of $[P]$.
\subsection{Modelling the resource coupling relationships}
\label{fmodel}
In this section, we discuss different models that we can use for the coupling functions. The first one is a linear regression modeled as $f^{t,t^\prime}_{v,v^{\prime}} (y,p) = \alpha^{t,t^\prime}_{v,v^{\prime}} y + \beta^{t,t^\prime}_{v,v^{\prime}} p + \gamma^{t,t^\prime}_{v,v^{\prime}} $. The parameters $\alpha^{t,t^\prime}, \beta^{t,t^\prime}, \gamma^{t,t^\prime} $ are obtained using the historical data collected in an offline manner. It is important to note that while we employ linear regression in this paper for modelling resource couplings, it is possible to use other models such as a support vector regressor (SVR) or a multilayer perceptron (MLP) resulting in better prediction performance. However, the benefit of linear regression models is that if the placement variables $x_{v,m}$ are assumed to be known, the resource allocation problem $[P]$ becomes a linear program (LP) for which efficient algorithms exist to generate the optimal solution in polynomial time. 
\begin{table*}[t]
\caption{\label{reg-perf} Performance comparison of linear, SVR and MLP models for resource coupling functions }
\centering
\begin{tabular}{c|c|c|c|c|c|c|c|c|c} 
 \hline
 Instance & \multicolumn{3}{c|}{\makecell{MAE \\  \hline Lin \quad  SVR \quad  MLP}} & \multicolumn{3}{c|}{\makecell{MSE \\ \hline Lin \quad SVR \quad  MLP}} & \multicolumn{3}{c}{\makecell{RMSE \\ \hline  Lin \quad  SVR \quad  MLP}} \\ [0.5ex]
  \hline\hline
 Exp 1-1, $f^{com,net}_{2,1}$ & 1.1 & 0.7 & 0.09& 2.6 &  3.2  & 0.01& 1.6  & 1.7 &  0.1\\
 \hline
 Exp 1-1, $f^{net,com}_{1,2}$  & 0.36 & 0.4& 0.17 & 0.17 &0.37 &0.04&  0.47 & 0.65 & 0.21\\
 \hline
 Exp 2-2, $f^{com,net}_{2,1}$ & 2.2 & 1.2 &0.4& 10.4 & 3.2  &0.4& 3.2 & 1.8 & 0.61\\
 \hline
 Exp 2-2, $f^{net,com}_{1,2}$ & 1.7  &1.17 &0.09& 6.4 &  5.6& 3.7& 2.5 &2.8 & 1.9 \\ [1ex] 
 \hline
 Exp 2-3, $f^{com,net}_{2,1}$ & 1.35 & 0.79 & 0.03& 4.16 & 5.43 &0.002& 2.04 & 2.33 & 0.053\\
 \hline
 Exp 2-3, $f^{net,com}_{1,2}$ & 2.05  &2.58 &0.57&7.94 & 15.71& 0.64& 2.81 & 3.96 & 0.80\\ [1ex] 
 \hline
\end{tabular}
\end{table*}
Table \ref{reg-perf} represents the performance of different regression models, for six coupling function examples of the watchlist and object detection applications, with coupling data shown in Fig. \eqref{fig:watchlist-com-br-lab}, \eqref{fig:com-br-lab} and \eqref{fig:com-br-car}. We use the polynomial kernel for the SVR model with $\gamma$ parameter of $ 10$ and the MLP has a hidden layer of size $100$ and uses $relu$ as activation function. It can be observed that the MLP regression model outperforms the SVR and linear regression models. However, the linear model is simple and useful for a special case of $[P]$ to become a LP as discussed earlier. In section \ref{sec:evaluation}, we illustrate that the usage of linear model for the coupling functions results in significant resource saving although it has limited prediction capability compared to SVR and MLP models.
In the rest of the paper, we consider a special case of $[P]$ with known placement decisions and linear models for the coupling functions, resulting in an LP. We demonstrate the benefits of incorporating the introduced resource coupling models in the application resource orchestration. For the placement problem, existing solutions such as the algorithm proposed in \cite{ECO} can be used accordingly.

\section{Performance Evaluation}
\label{sec:evaluation}
In this section, we present the experimental setup and benchmark our proposed solution, \system, against a static resource allocation scheme, which ignores the coupling between resources. 
\subsection{Experiment setup}
Our experimental setup is shown in Fig. \ref{fig-testbed-architecture}, where we have wireless gateways from Multitech \cite{multitech-link} and Access Point (AP) from Celona \cite{celona-link}. User Equipment (UE) connect over private 5G to AP. 5G core and MEC servers are in our internal LAN and the core is remotely configured using Celona's Service Orchestrator. Control and Data plane traffic from AP is terminated at the core. In our MEC setup, we have one master and three worker node servers. Master node is equipped with 10-core Intel core i9 CPU and the three worker nodes are equipped with 24-core Intel CPU and with NVIDIA RTX 2080 Ti GPUs. Kubernetes \cite{kubernetes-link} cluster is setup on our MEC servers and both our usecases i.e. video surveillance (watchlist application) and intelligent transportation systems (object detection application) run within pods in Kubernetes. Each function runs as a separate pod and multiple replicas of these pods are created, as necessary. We stream videos from video server using ffmpeg \cite{ffmpeg-link} and they are processed in MEC servers on a Kubernetes cluster, within pods. We use GNU Linear Programming Kit (GLPK) \cite{glpk} solver for the optimization problem discussed in section \ref{sec:solution}.
\begin{figure}[b]
    \centering
    \includegraphics[width=0.96\linewidth]{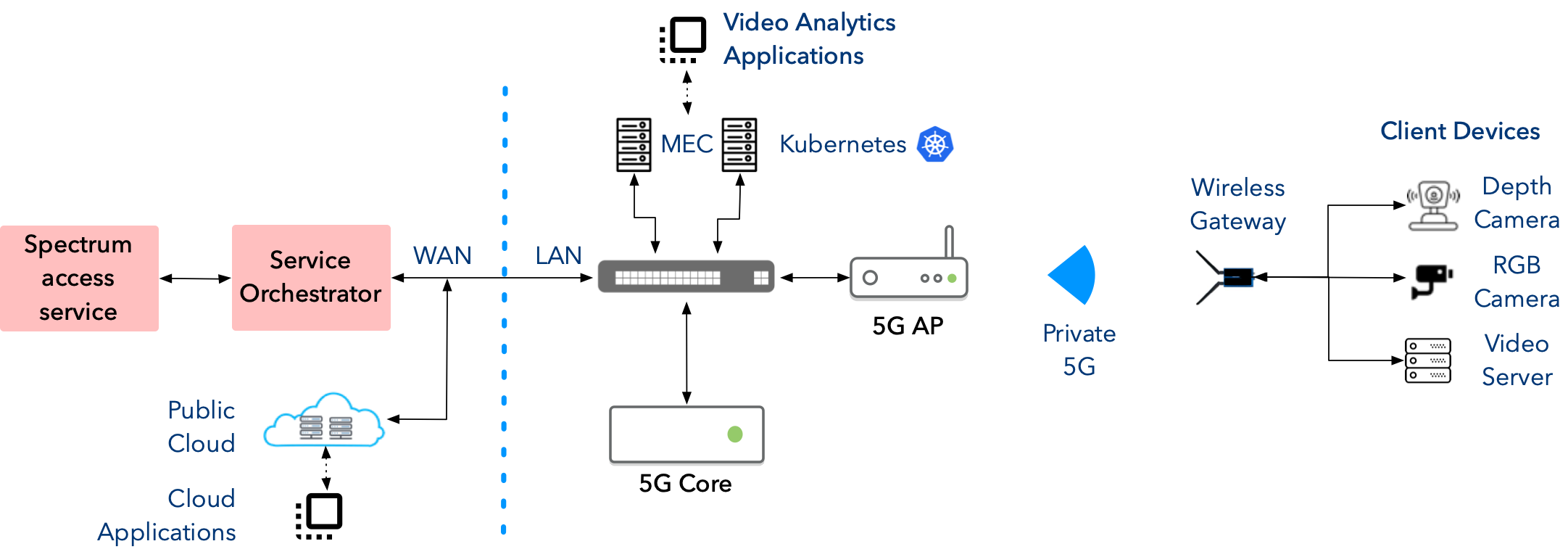}
    \caption{Experimental setup}
    \label{fig-testbed-architecture}
\end{figure}
\subsection{Results}
\begin{figure*}[t] 
\centering
  \begin{subfigure}{0.23\textwidth}
    \includegraphics[width=\linewidth]{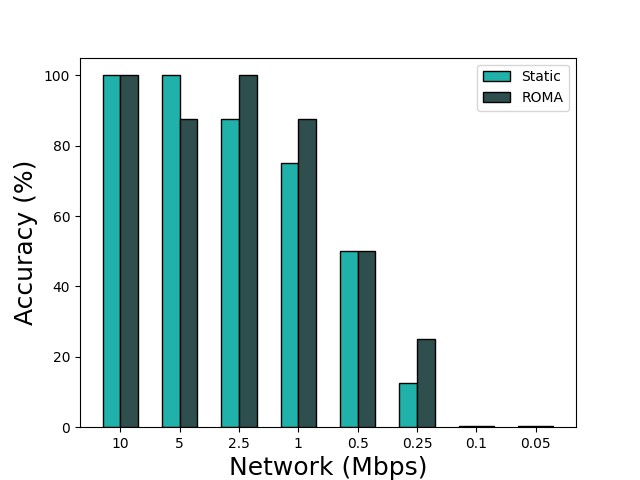}
    \caption{Application accuracy}
    \label{fig:performance-watchlist}
  \end{subfigure}
  \begin{subfigure}{0.23\textwidth}
    \includegraphics[width=\linewidth]{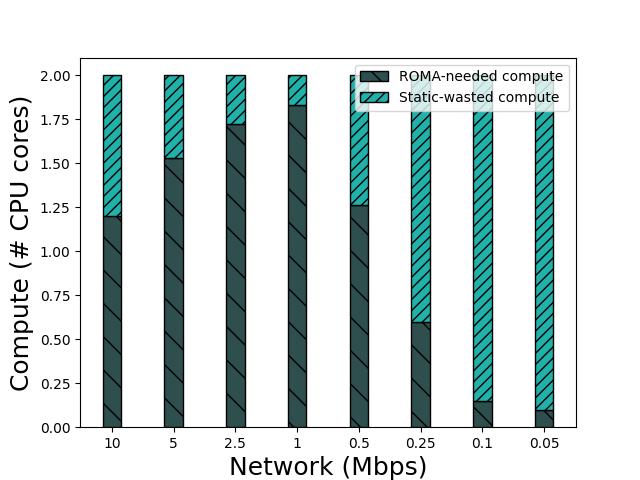}
    \caption{Resource usage}
    \label{fig:resource-watchlist}
  \end{subfigure}
     \begin{subfigure}{0.23\textwidth}
    \includegraphics[width=\linewidth]{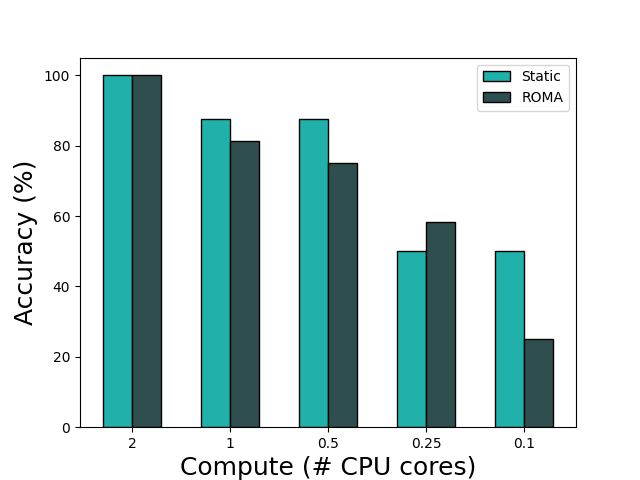}
    \caption{Application accuracy}
    \label{fig:performance-watchlist2}
  \end{subfigure}
  \begin{subfigure}{0.23\textwidth}
    \includegraphics[width=\linewidth]{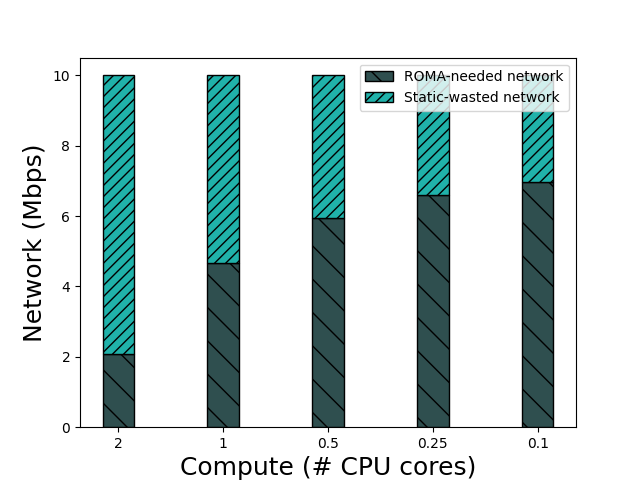}
    \caption{Resource usage}
    \label{fig:resource-watchlist-network}
  \end{subfigure}
  \label{fig:watchlist-results}
  \caption{Performance of watchlist application, \system\ vs. static}
\end{figure*}  
\begin{figure*}[t] 
\centering
  \begin{subfigure}{0.23\textwidth}
    \includegraphics[width=\linewidth]{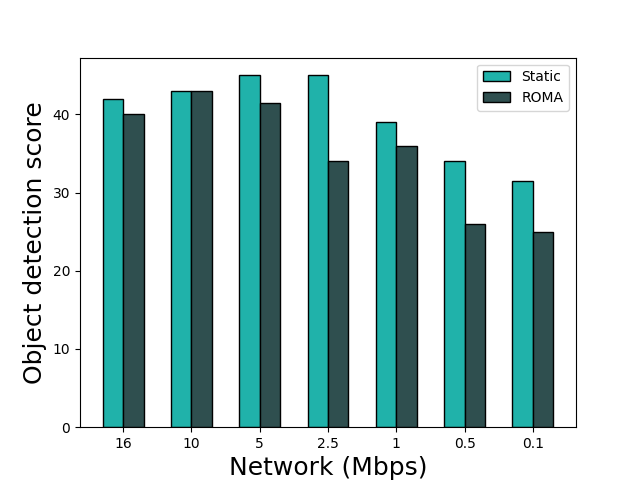}
    \caption{Detection score}
    \label{fig:performance-obj}
  \end{subfigure}
  \begin{subfigure}{0.23\textwidth}
    \includegraphics[width=\linewidth]{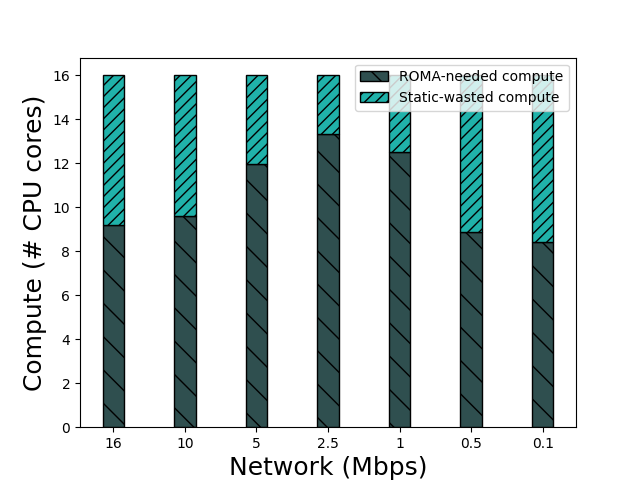}
    \caption{Resource usage}
    \label{fig:resource-obj}
  \end{subfigure}
    \begin{subfigure}{0.23\textwidth}
    \includegraphics[width=\linewidth]{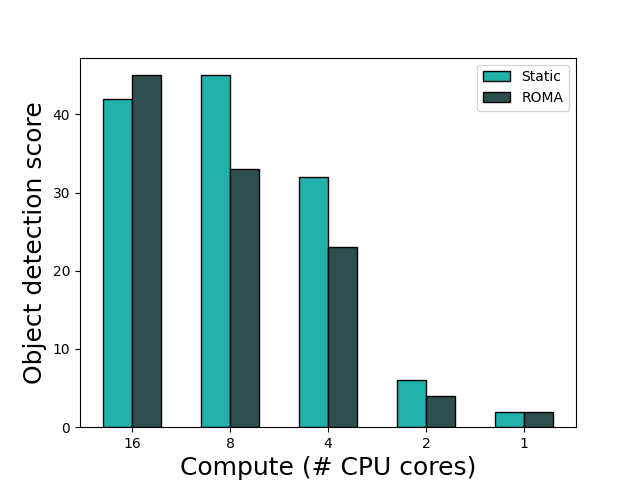}
    \caption{Detection score}
    \label{fig:performance-obj2}
  \end{subfigure}
  \begin{subfigure}{0.23\textwidth}
    \includegraphics[width=\linewidth]{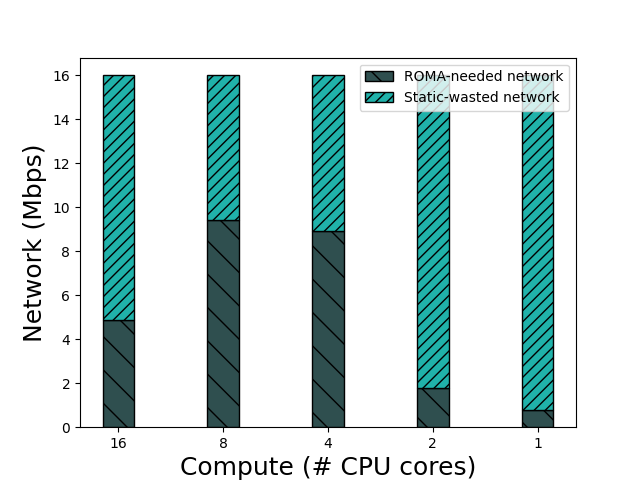}
    \caption{Resource usage}
    \label{fig:resource-obj-network}
  \end{subfigure}
\caption{Performance of object (person) detection application: \system\ vs. static} \label{fig:person-detection-results}
\end{figure*}
\begin{figure*}[t] 
\centering
  \begin{subfigure}{0.22\textwidth}
    \includegraphics[width=\linewidth]{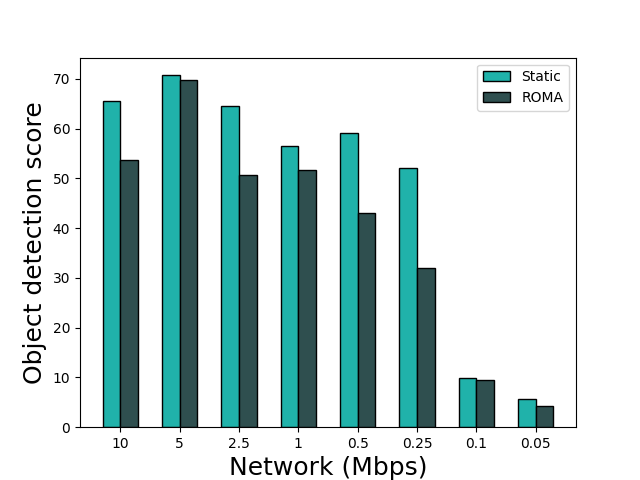}
    \caption{Detection score}
    \label{fig:performance-car}
  \end{subfigure}
  \begin{subfigure}{0.23\textwidth}
    \includegraphics[width=\linewidth]{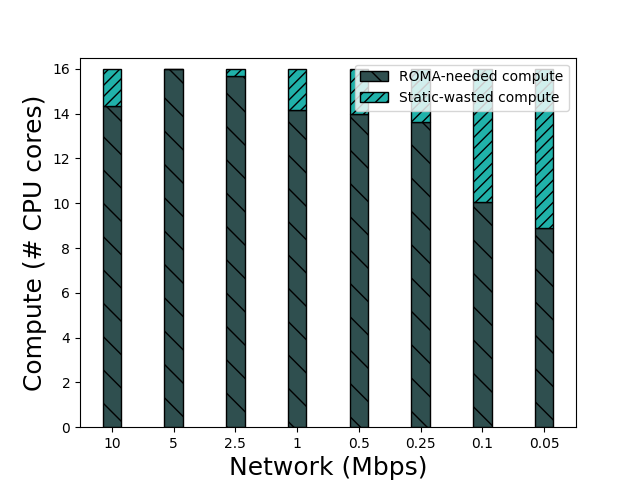}
    \caption{Resource usage}
    \label{fig:resource-car}
  \end{subfigure}
    \begin{subfigure}{0.23\textwidth}
    \includegraphics[width=\linewidth]{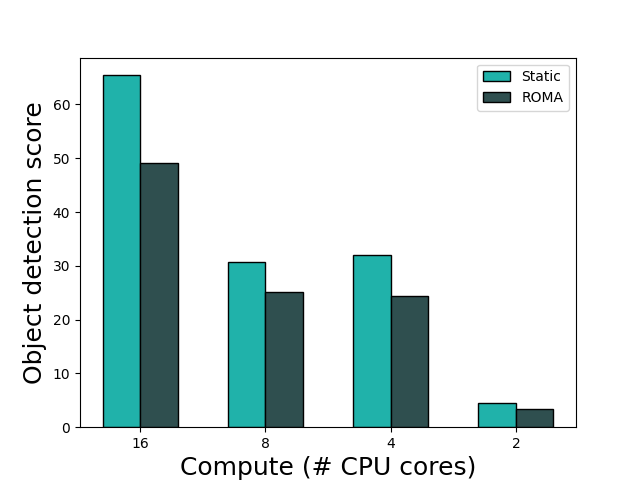}
    \caption{Detection score}
    \label{fig:performance-car2}
  \end{subfigure}
  \begin{subfigure}{0.23\textwidth}
    \includegraphics[width=\linewidth]{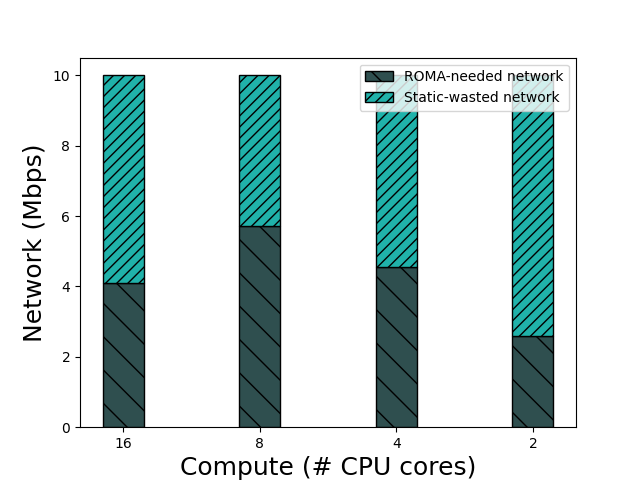}
    \caption{Resource usage}
    \label{fig:resource-car-network}
  \end{subfigure}
\caption{Performance of object (car) detection application: \system\ vs. static} \label{fig:car-detection-results}
\end{figure*}
We present our experimental results for the two use cases i.e. video surveillance and intelligent transportation system:
\subsubsection{Video surveillance}
For the video surveillance use case, we deploy the watchlist application depicted in Fig. \ref{fig:watchlist service}. We consider a sample video including different people and compare the performance of two resource allocation strategies, \system\, and a static allocation in which the amount of compute allocated to the face detection function is fixed  to $2$ cores. The results are averaged over $3$ runs.  Fig. \eqref{fig:performance-watchlist} shows the average application accuracy as the network experiences congestion and the bandwidth drops from $10\ Mbps$ to $0.05\ Mbps$. In Fig. \eqref{fig:resource-watchlist}, the compute resource usage is compared for the two schemes as the network bandwidth changes. It is observed that compared to the static resource allocation scheme, \system\ is able to reduce the compute resource usage up to $90\%$, hence preventing overprovisioning while maintaining the application accuracy, by exploiting the network-compute coupling relationship. It is important to note that the slight variation in the application accuracies in Fig. \eqref{fig:performance-watchlist} is mainly due to the fact that different set of frames may be processed in each case because some frames are dropped by the video ingress function until CPU is released to process next frame, and that is why we need to average the results over multiple runs. Similar results are shown in Fig. \eqref{fig:performance-watchlist2} and \eqref{fig:resource-watchlist-network} when the compute resource changes, and the application accuracy and network resource usage is compared for \system\ and the static scheme. In particular, according to Fig \eqref{fig:resource-watchlist-network}, \system\ reduces the bandwidth up to $80\%$ while the application accuracy remains acceptable. 


\subsubsection{ITS} In this experiment, we implement the object detection application shown in Fig. \ref{fig:object detecting service}. We consider the used videos in Exp 2-2 and Exp 2-3 in Fig \eqref{fig:com-br-lab} and \eqref{fig:com-br-car}. The goal is to detect the person or car objects in the videos. Fig. \eqref{fig:performance-obj} demonstrates the application performance in terms of the object detection score, which we define next. The amount of used compute resource, as the network condition changes is also showed in Fig \eqref{fig:resource-obj}. Since not all the video frames are processed at each experiment instance, and because the number of objects differ in various video frames, we define the following weighted score (which is different from confidence score) for the object detection application:
\begin{align}
    score = \sum_{f \in FRAME} w_{f} \frac{TP_{f}}{GT_{f}}
\end{align}
where $f, TP_{f} $ and $GT_{f}$ denote the frame index, the number of true positives in frame $f$ and the number of ground truth objects in frame $f$. Moreover, $FRAME$ denotes the set of processed frames. We use the intersection over union (IoU) metric to measure the overlap between the detected and ground truth bounding boxes. The IoU threshold is predefined as $0.5$ and the predictions with an IoU of $0.5$ and above are classified as TP.
Fig. \eqref{fig:performance-obj} and \eqref{fig:resource-obj} illustrate that \system\ trades off a small degree of performance for significant compute resource saving by leveraging the network-compute coupling relationship in the resource allocation. The amount of compute resource usage is reduced up to $50\%$ in this case. In Fig. \eqref{fig:performance-obj2} and \eqref{fig:resource-obj-network}, the object detection score and the network resource usage are shown respectively as the available compute resource varies. It is observed that \system\ outperforms the static scheme in terms of compute resource usage by saving up to $95\%$ of the network bandwidth (in the case that the available number of CPU cores is $1$), while the object detection score is comparable with the overprovisioned static solution. The same results for the car detection application are depicted in Fig. \ref{fig:car-detection-results}. According to Fig. \eqref{fig:performance-car} and \eqref{fig:resource-car}, as the network bandwidth varies, \system\ is able to save on the compute resource usage up to $44\%$ compared to the static approach. The detection score remains within an acceptable range of the static solution except for the single case of $0.25\ cores$. From Fig. \eqref{fig:performance-car2} and \eqref{fig:resource-car-network}, as the compute resource changes from $16\ cores$ to $2\ cores$ for the car detection experiment, the network bandwidth is remarkably saved up to $75\%$ in the case with $2\ cores$.

\section{Related Work}
\label{sec:related}
The work presented in \cite{liu2019direct}, uses a distributed cross-domain resource orchestration (DIRECT) for cellular edge computing considering the radio and computing resources of a radio access network and multiple edge servers. The formulated resource orchestration problem takes the perspective of the network operator with the objective of maximizing the sum utility of network slices on all edge nodes. Assuming unknown utility functions, the authors propose a learning-assisted resource orchestration based on a probabilistic model with a gradient-based optimization solution. Authors in \cite{Sl-EDGE} show that DIRECT incurs in overprovisioning due to ignoring the coupling between different edge resources. This coupling is modeled as a linear function used in an MILP to optimally instantiate joint network-MEC slices and prevent resource overprovisioning. In order to deal with the MILP time complexity, distributed algorithms are proposed to leverage the similarities among edge nodes and resource virtualization which can instantiate heterogeneous slices within a short distance from the optimum. In the literature, there are also studies such as \cite{8768602,9289885} focusing on the problem of virtual network embedding which shares similarities with the microservices-based application deployment problem. The main limitation of these works is that they only consider the edge or core networks which cannot be readily applied to a multi-tier computing infrastructure. 

In \cite{sonmez2020machine}, a multi-tier vehicular edge computing (VEC) system is considered which consists of three layers of data generation, vehicular edge computing and remote cloud. The overall system involves the cooperation of local edge servers with the global cloud servers distributed over a geographical region. An ML-based prediction is utilized to performs a two-stage process for the offloading decision, a classification for predicting the offloading success and a regression for the service time estimation.
While the solution in\cite{sonmez2020machine} takes into account static dataset and models, authors in \cite{alperformance} propose online multi-armed bandit (MAB) based task offloading schemes to avoid poor performance when the VEC environment conditions to which the static models are exposed differ from  those  used  for model training. It is shown that the proposed contextual bandit-based algorithm surpass all other algorithms under the failure rate and QoE metrics besides achieving adequately comparable service time values. Different from \cite{sonmez2020machine} and \cite{alperformance}, we consider an underlying realistic 5G system. Moreover, the workloads in the aforementioned works are only single tasks, which makes the proposed frameworks not applicable to the microservices-based applications. Authors in \cite{eXP-RAN} developed an emulation software named eXP-RAN, which allows experimenting with network slicing in virtualized RAN nodes and EC scenarios, with the key characteristics of slicing abstraction, service representation, and predictable performance. Compared with existing relevant tools such as 
EdgeCloudSim \cite{sonmez2018edgecloudsim}, eXP-RAN has the ability of monitoring each network slice independently. However, current version of eXP-RAN only implements the RAN slicing and it lacks 5G core network implementation. Therefore, its applicability for a case of multi-tier computing framework is questionable. Regarding the microservices-based application deployment in a dynamic environment, authors in \cite{ray2020proactive} design an RL-based proactive scheme for placement and migration of an already placed microservice in the MEC setup. In contrast to a conservative policy leading to wasteful resource allocation and a reactive on-demand policy causing high latency, the main contribution of this paper is to learn and synthesize the optimal proactive prefetch, deployment and migration schedule, given a microservice workflow by utilizing the user mobility. However, only sequential workflow structure (linear sequence of microservices) is considered in this study and no back-end central cloud is assumed. Moreover, all the microservices invoked by a user can be co-located at one edge server.

\section{Conclusion}
\label{sec:conclusion}
Emerging IoT applications today have strict and diverse set of requirements in terms of end-to-end latency, throughput, reliability, etc. 5G and edge computing has made it possible to meet these requirements. However, with the addition of extra computing and networking layers, the underlying infrastructure itself has become quite complex. To this end, in this paper, we propose \system, which enables optimal resource orchestration for microservices-based 5G applications in a dynamic, heterogeneous, multi-tiered compute and network fabric by exploiting resource coupling relationships. \system\ is able to successfully map application-level requirements to individual functions and ensure optimal deployment, such that end-to-end application requirements are met. By implementing two real-world IoT applications, one in video surveillance domain and another one in intelligent transportation systems domain, we show that \system\ is able to reduce compute and network resource usage remarkably while maintaining the application performance. In future, we aim at studying the performance of  machine learning-based dynamic resource allocation which potentially captures nonlinear resource coupling relationships.

\bibliographystyle{IEEEtran}
\bibliography{ref}
\end{document}